\documentclass[apj,square,numbers]{emulateapj}
\usepackage{graphicx}
\usepackage{xcolor}
\usepackage{amsmath,amssymb,latexsym}
\usepackage{hyperref}
\usepackage{ulem}
\begin{document}
\title{Model-independent estimations for the curvature from standard candles and clocks}

\author{Zhengxiang Li$^{1,2}$, Guo-Jian Wang$^{1,2}$, Kai Liao$^{3}$, and Zong-Hong Zhu$^{1,2}$}

\affil{$^{1}$Department of Astronomy, Beijing Normal University, Beijing 100875, China}
\affil{$^{2}$Laboratory of Gravitational Waves and Cosmology, Beijing Normal University, Beijing 100875, China}
\affil{$^{3}$School of Science, Wuhan University of Technology, Wuhan 430070, China}

\begin{abstract}

  Model-independent estimations for the spatial curvature not only provide a test for the fundamental Copernican principle assumption, but also can effectively break the degeneracy between curvature and dark energy properties. In this paper, we propose to achieve model-independent constraints on the spatial curvature from observations of standard candles and standard clocks, without assuming any fiducial cosmology and other priors. We find that, for the popular Union2.1 type Ia supernovae (SNe Ia ) observations, the spatial curvature is constrained to be $\Omega_K=-0.045_{-0.172}^{+0.176}$. For the latest joint light-curve analysis (JLA) of SNe Ia observations, we obtain $\Omega_K=-0.140_{-0.158}^{+0.161}$. It is suggested that these results are in excellent agreement with the spatially flat Universe. Moreover, compared to other approaches aiming for model-independent estimations of spatial curvature, this method also achieves constraints with competitive precision.

\end{abstract}


\maketitle

\section{Introduction}\label{sec:intro}
The spatial curvature of the Universe is one of the most fundamental issues in modern cosmology. Specifically, on one hand, estimating the curvature
of the Universe is a robust way to test the important assumption that the Universe is exactly described by the homogeneous and isotropic Friedmann-Lema\^{i}tre-Robertson-Walker (FLRW) metric. On the other hand, the curvature of the Universe is also closely related to some other
important problems such as the evolution of the Universe and the nature of dark energy. For instance, nonzero curvature may result in enormous effects on reconstructing the state equation of dark energy even though the true curvature might be very small \citep{Ichikawa2006,Clarkson2007,Gong2007,Virey2008}, and any significant deviation from the flat case would lead to profound consequences for inflation models and fundamental physics. Moreover, possibilities for the failure of the FLRW
approximation have been proposed to account for the observed late-time accelerated expansion \citep{Ferrer2006,Ferrer2009,Enqvist2008,Redlich2014,Rasanen2009,Lavinto2013,Boehm2013}. Therefore, observational constraints on the cosmic curvature from popular probes have been extensively studied in the literature \citep{Eisenstein2005,Tegmark2006,Zhao2007,Wright2007}. We emphasize here that, a spatially flat Universe in the framework of the standard $\Lambda$CDM model is favored at very high confidence level
by the latest {\it Planck} 2015 results of Cosmic Microwave Background (CMB) observations \citep{Planck2015}. However, all these works did
not measure the curvature in any direct geometric way. That is, curvature is primarily derived from a measurement of the angular diameter distance to recombination, which not only depends on curvature but also on the choice of cosmological model assumed in the analysis.

Recently, \citet{Clarkson2008} proposed to measure the spatial curvature of the Universe or even test the FLRW metric in a model-independent way by combining observations of expansion rate and distance, which has been fully implemented with updated observational data~\citep{Shafieloo2010, Mortsell2011, Sapone2014, Ronggen2016}. In this method, derivative of distance with respect to redshift $z$ is necessary to estimate the
curvature, this treatment introduces a large uncertainty. Therefore, \citet{Yu2016} improved this method by confronting the distances derived from expansion rate and Baryon Acoustic Oscillations (BAO)  observations. However, the angular diameter distance data and some expansion rate measures used in their analysis are obtained from BAO observations, which are dependent on the assumed fiducial cosmological model and the prior for the distance to the last-scattering surface from cosmic microwave background (CMB) measurements. Another method was also put forward to attain a similar test by using parallax distances and angular diameter distances~\citep{Rasanen2014}. In addition, the cross-correlation between foreground mass and gravitational shear of background galaxies has been proposed to be a practical measurement of the curvature of the Universe, which purely relies on the properties of the FLRW metric~\citep{Bernstein2006}. More recently, the sum rule of distances along null geodesics of the FLRW metric has been put forth as a consistency test~\citep{Rasanen2015}. It is interesting to note that, on one hand, the FLRW background will be ruled out if the sum rule is violated; on the other hand, if the observational data is well consistent with the sum rule, the test provides a model-independent estimation of the spatial curvature of the Universe. In their analysis, by using the Union2.1 compilation of type Ia supernova (SNe Ia)~\citep{Suzuki2012} and strong gravitational lensing data selected from the Sloan Lens ACS Survey~\citep{Bolton2008}, the spatial curvature parameter was weakly constrained to be $\Omega_K=-0.55_{-0.67}^{+1.18}$ at $95\%$ confidence level, which slightly favors a spatial closed Universe. Actually, the distances used in their analysis from the Union2.1 SN Ia are not completely cosmology-free, since the light-curve fitting parameters accounting for distance estimation are determined from a global fit in the assumed standard dark energy model with the equation of state being constant. Although such an effect is likely subdominant to the uncertainties, the measurement of spatial curvature was not so cosmological-model-independent as they claimed. Moreover, for the distance sum rule (the Eq. 4 in \citet{Rasanen2015}) used to calculate the spatial curvature, it is argued that this formula is only valid for the case of $\Omega_K\geq0$~\citep{Hogg99}.

In this paper, firstly, we reconstruct a function of Hubble parameter with respect to redshift $z$ from expansion rate measures of cosmic chronometers (or standard clocks) by using a non-parametric smoothing method (\citet{Li2016}: NPS hereafter). This reconstructed function enablse us to directly get the comoving distance by calculating the integral of it. Next, with the spatial curvature parameter taken into consideration, we transform these comoving distances into curvature-dependent luminosity distances. Then, by confronting them with luminosity distances depending on light-curve fitting parameters from SNe Ia observations, we achieve cosmological model-independent constraints on the spatial curvature. For the Union2.1 SN Ia, we obtain $\Omega_K=-0.045_{-0.172}^{+0.176}$. When the latest JLA SN Ia~\citep{Betoule2014} is used, the spatial curvature is constrained to be $\Omega_K=-0.140_{-0.158}^{+0.161}$. On one hand, these results consistently favor a spatially flat Universe. On the other hand, in the context of model-independent estimations for spatial curvature, these constraints are comparable in terms of  precision.

\section{Method and Data}\label{sec:method}

\subsection{Distance from expansion rate measurements}
The expansion rate at any redshifts $z\neq0$, $H=\dot{a}/{a}$ where $a = 1/(1+z)$, can be obtained from the derivative of redshift with respect to cosmic time, i.e., $H(z)\simeq-\frac{1}{1+z}\frac{\Delta z}{\Delta t}$. The difficulty of this approach is to estimate the change in the age of the Universe as a
function of redshift $\Delta t$. \citet{Jimenez02} proposed to make this method practicable by calculating the age difference between two luminous red galaxies at different redshifts. In the literature, this method is usually referred to as differential age (DA) and the passively evolving galaxies from which $\Delta t$ is estimated are called cosmic chronometers. So far, 22 measurements of $H(z)$ based on this method  (in the redshift range $0.070 \leq z \leq 1.965$) have been obtained~\citep{Jimenez03,Simon05,Stern10,Moresco12,Moresco:2015cya,Moresco2016}. Although cosmological model-independent, some of these estimates are sensitive to stellar population synthesis models whose influence on $\Delta t$. It was found that this influence becomes important at $z \gtrsim 1.2$~\citep{Verde2014}. Therefore, we consider only 16 $H(z)$ measurements in the range $z < 1.2$ which, in practice, given the redshift distribution of the $H(z)$ data, means $z \leq 1.037$. In addition, we also slightly increase (20\%) the error bar of the highest-$z$ point to account for the uncertainties of the stellar population synthesis models~\citep{Verde2014}. This ensures that the evolution of the reconstructed Hubble parameter as a function of redshift in the following analysis is neither dependent on the cosmology nor on the stellar population synthesis model.

\begin{figure*}[t]
	\centering
	\includegraphics[width=0.98\textwidth, height=0.32\textwidth]{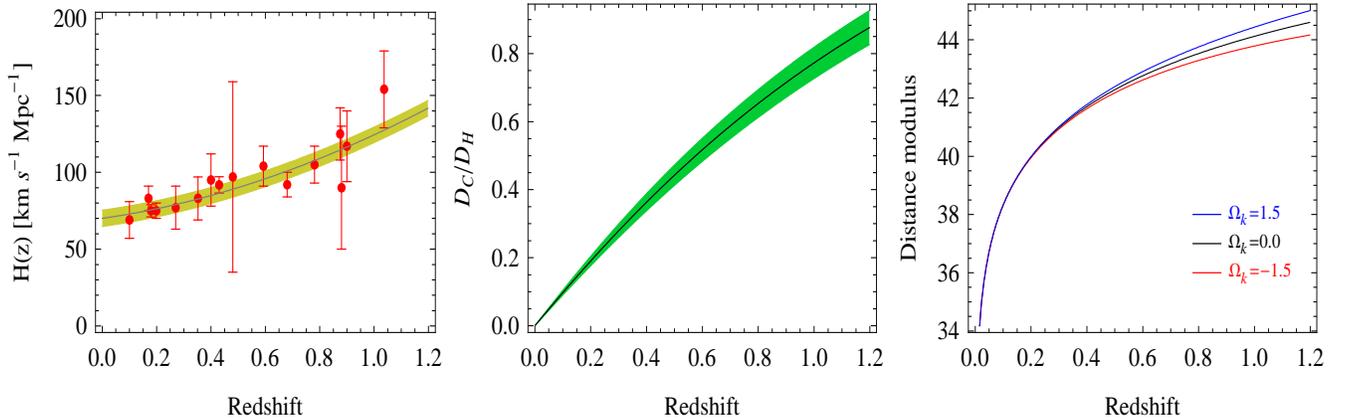}\\
	\caption{{\bf Left:} Measured expansion rates from cosmic chronometers observations and the function of Hubble parameter versus redshift reconstructed from the observational data with the NPS method. {\bf Middle:} The corresponding reconstructed $D_\mathrm{C}/D_\mathrm{H}$. {\bf  Right:} Indicative dependence of the distance-redshift relation derived from the $D_\mathrm{C}/D_\mathrm{H}$ (Eq.~\ref{eq1}) on the spatial curvature $\Omega_K$.}\label{fig1}
\end{figure*}

In~\citet{Li2016}, we reconstructed a reasonable function of Hubble parameter versus redshift with the NPS method, which is an improved version of the smoothing process proposed by~\citet{Shafieloo06}. The 16 expansion rate measurements of cosmic chronometers and the reconstructed function with  $1\sigma$ confidence region are shown in the left panel of Fig.~\ref{fig1}. As proposed in~\citet{Busti2014}, by extrapolating this function to redshift $z=0$, we can obtain the model-independent determination of  the Hubble constant, $H_0$, from intermediate redshift cosmic chronometers observations. In their analysis, using the Gaussian processes~\citep{Seikel2012}, they obtained a lower $H_0$ than the CMB value and hence reinforced the tension with the local measurement. Here, we find $H_0=69.407\pm6.031$,  which is in an agreement with both the latest CMB value~\citep{Planck2015} and local measurement~\citep{Riess2016}. For consistency, this extrapolated $H_0$ is also used for distance estimation in the following analysis. Enlightened by the method which roughly transforms discrete $H(z)$ measurements into comoving distances by solving the integral with a simple trapezoidal rule~\citep{Holanda2013}, we obtain the comoving distances at $z < 1.2$ by integrating the smoothed function of Hubble parameter with respect to redshift.  It is known that the comoving distance $D_\mathrm{C}$ connects the luminosity distance $D_\mathrm{L}$ via~\citep{Hogg99},
\begin{equation}\label{eq1}
\frac{D_\mathrm{L}}{(1+z)}=\begin{cases} \frac{D_\mathrm{H}}{\sqrt{\Omega_K}} \sinh{[\sqrt{\Omega_K}D_\mathrm{C}/D_\mathrm{H}]}&\Omega_K>0\\D_\mathrm{C} &\Omega_K=0\\ \frac{D_\mathrm{H}}{\sqrt{\left| \Omega_K \right|}} \sin{[\sqrt{\left| \Omega_K \right|}D_\mathrm{C}/D_\mathrm{H}]}&\Omega_K<0,
\end{cases}
\end{equation}
where $D_\mathrm{H}=cH_0^{-1}$ and $c$ is the speed of light. With the extrapolated $H_0$, we obtain the curvature-free $D_\mathrm{C}/D_\mathrm{H}$ and the result is presented in the middle panel of Fig.~\ref{fig1}. Moreover, in the right panel of Fig.~\ref{fig1}, 
we illustrate the dependence of  distance modulus, $\mu=5\log\big[\frac{D_\mathrm{L}}{\mathrm{Mpc}}\big]+25$, derived from cosmic chronometers observations on the spatial curvature. 

\subsection{Distance from type Ia supernovae observations}
The distance estimation from SNe Ia data is on the basis of the empirical observation that these events form a homogeneous class whose remaining variability is reasonably well captured by two parameters. One of them describes the time stretching of the light-curve ($x_1$) whereas the other  describes the SNe Ia color at maximum brightness ($c$). For the popular Union2.1 SN Ia~\citep{Suzuki2012}, where the SALT2 model is used to reconstruct light-curve parameters ($x_1$, $c$, and the observed peak magnitude in rest frame $B$ band $m_\mathrm{B}^*$),  the distance estimator assumes that SNe Ia with identical color, shape, and galactic environment have on average the same intrinsic luminosity at all redshifts. This assumption can yield a linear expression to standardize the distance modulus,
\begin{equation}\label{eq2}
\begin{split}
\mu^{\mathrm{SN}}(\alpha, \beta, \delta, M_\mathrm{B})=m_\mathrm{B}^*-M_\mathrm{B}+\alpha\times x_1-\beta\times c\\+\delta \cdot P(m^{\mathrm{true}}_*<m_*^{\mathrm{threshold}}),
\end{split}
\end{equation}
where $\alpha$ and $\beta$ are nuisance parameters which characterize the stretch-luminosity and color-luminosity relationships, reflecting the well-known broader-brighter and bluer-brighter relationships, respectively. The value of $M_\mathrm{B}$ is another nuisance parameter which represents the
absolute magnitude of a fiducial SNe. In addition, the term $\delta \cdot P(m^{\mathrm{true}}_*<m_*^{\mathrm{threshold}})$ with $m_*^\mathrm{threshold}=10^{10}m_\odot$ is introduced to account for the host-mass correction to SNe Ia luminosities~\citep{Sullivan2010}. 

In the latest JLA SN Ia~\citep{Betoule2014}, light-curve parameters are also obtained with the SALT2 model and the distance modulus is estimated by a similar expression of the Eq.~\ref{eq2} but without the term of host-mass correction. Alternatively, they approximately correct for the effect of dependence of absolute magnitude $M_\mathrm{B}$ on properties of host galaxies, e.g., the host stellar mass ($M_{\mathrm{stellar}}$), with a simple step function when the mechanism is not fully understood~\citep{Sullivan2011,Conley2011},
\begin{eqnarray}\label{eq3}
M_\mathrm{B}=\begin{cases} M_\mathrm{B}^1~~~&\mathrm{if}~
M_{\mathrm{stellar}}<10^{10} M_\odot.\\
M_\mathrm{B}^1+\Delta_\mathrm{M}~~&\mathrm{otherwise}.
\end{cases}
\end{eqnarray}

In general, the light-curve fitting parameters, $\alpha$ and $\beta$, and $\delta$ are left as free parameters being determined in the global fit to the Hubble diagram in the framework of the standard dark energy model. This treatment results in the dependence of distance estimation on the cosmological model used in the analysis. Therefore, implications derived from SNe Ia observations with the light-curve fitting parameters determined in the global fit to the Hubble diagram are somewhat cosmological-model-dependent.

\section{Results}
In order to achieve model-independent estimation for the cosmic curvature, rather than using the distance modulus versus redshift data published in  conventional SNe Ia samples, we confront light-curve fitting parameters-dependent distances in Eq.~\ref{eq2} with curvature-dependent ones from cosmic chronometers observations in Eq.~\ref{eq1} by maximizing the following likelihood:
\begin{equation}
\mathcal{L}(\Omega_K; \textbf{P}_{\mathrm{SN}})\propto\prod_{i=1}^{\mathrm{SNe}}\exp\bigg[-\frac{(\mu^{\mathrm{CC}}(z_i; \Omega_K)-\mu^{\mathrm{SN}}(z_i; \textbf{P}_{\mathrm{SN}}))^2}{2(\sigma^2_{\mu^{\mathrm{CC}}}+\sigma^2_{\mu^\mathrm{SN}})}\bigg],
\end{equation}
where $\textbf{P}_{\mathrm{SN}}$ stands for parameters in SNe Ia distance estimation, including light-curve fitting parameters ($\alpha$, $\beta$) and paremeters connecting to the intrinsic luminosity ($M_\mathrm{B}$ and $\delta$, or $M^1_{\mathrm{B}}$ and $\Delta_\mathrm{M}$), $\sigma^2_{\mu^\mathrm{SN}}$ accounts for error in SNe Ia observations propagated from the covariance matrix~\citep{Amanullah2010, Conley2011}. In our analysis,  we use the covariance matrix with both the reported statistical and systematic errors.  Then, we use emcee\footnote{https://pypi.python.org/pypi/emcee} introduced by~\citet{Foreman2012}, a Python module that includes Markov chain Monte Carlo (MCMC), to get the best-fit values and their corresponding uncertainties for both $\Omega_K$ and  parameters in SNe Ia distance estimation by generating sample points of the probability distribution.

\begin{figure}[t]
	\centering
	\includegraphics[width=0.47\textwidth, height=0.47\textwidth]{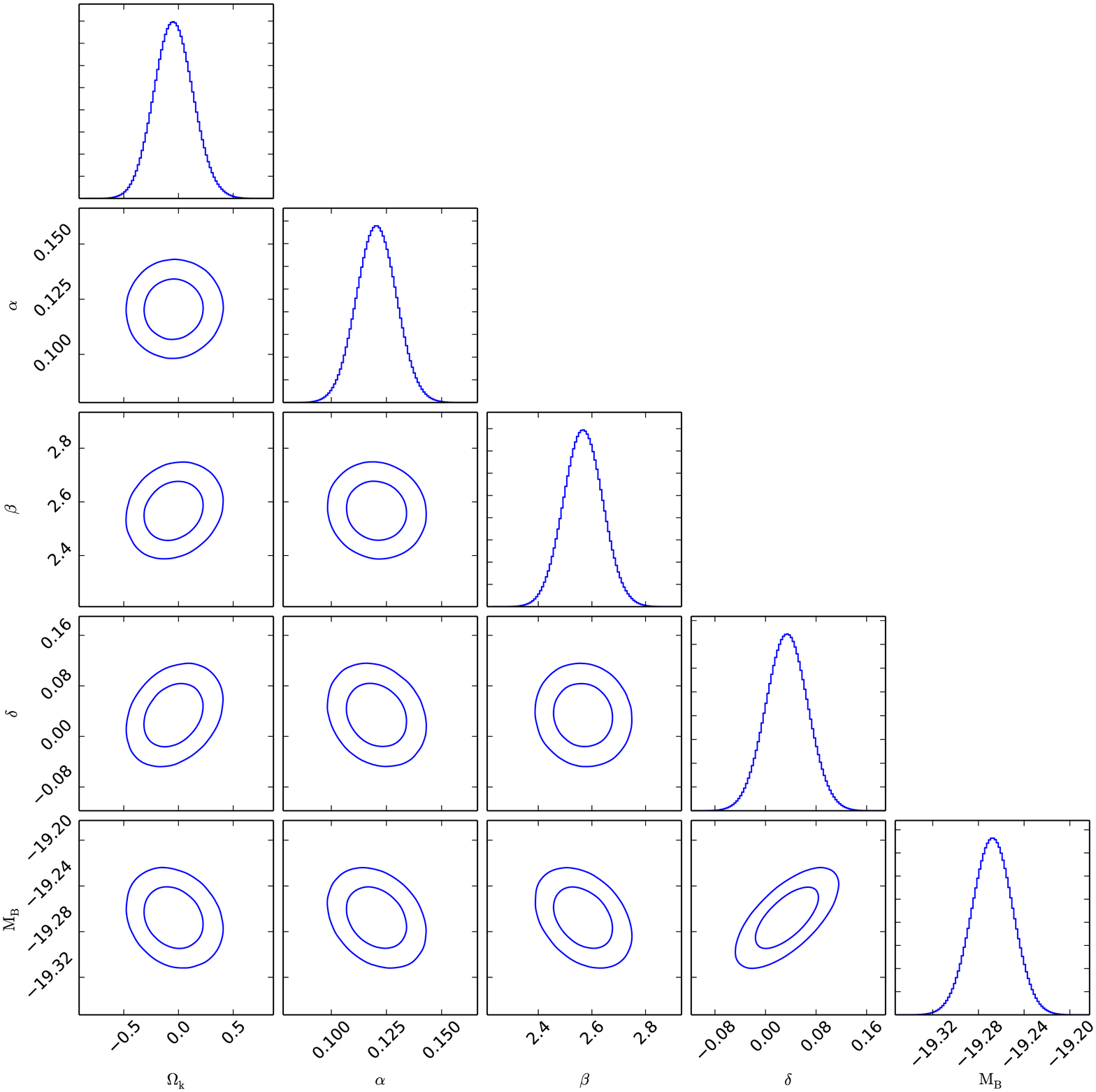}\\
	\caption{68\% and 95\% confidence level contours for the spatial curvature and light-curve fitting parameters model-independently constrained from the Union2.1 SNe Ia and cosmic chronometers observations. }\label{fig2}
\end{figure}

\begin{figure}[t]
	\centering
	\includegraphics[width=0.47\textwidth, height=0.47\textwidth]{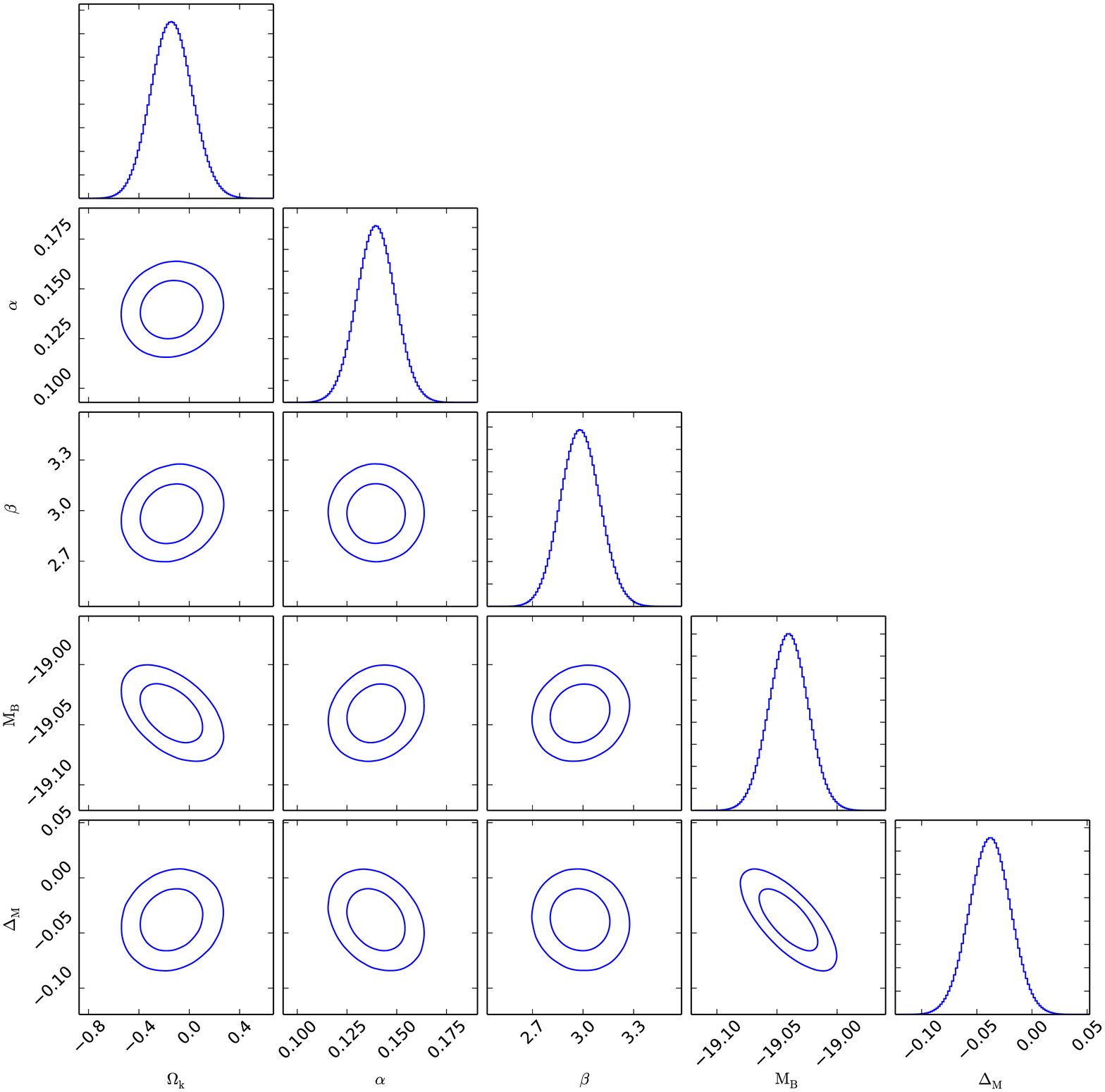}\\
	\caption{68\% and 95\% confidence level contours for the spatial curvature and light-curve fitting parameters model-independently constrained from the JLA SNe Ia and cosmic chronometers observations. }\label{fig3}
\end{figure}

For the Union2.1 SN Ia, 563 well-measured SNe Ia events remain for the likelihood estimation because of the redshift cutoff $z<1.2$ for distance derived from expansion rate measurements. Results are shown in Fig.~\ref{fig2} and summarized in Tab.~\ref{tab1}. We find that, from the Union2.1 SNe Ia and cosmic chronometers observations, model-independent estimation for the spatial curvature is $\Omega_K=-0.045_{-0.172}^{+0.176}$. This is in full agreement with the constraints obtained from the latest $Planck$ CMB measurements~\citep{Planck2015}. Moreover, the precision of this estimation is more competitive than the model-independent test based on the distance sum rule~\citep{Rasanen2015}. When the JLA SN Ia is used, 737 well-measured events distribute in the range $z<1.2$.  Results are shown in Fig.~\ref{fig3} and summarized in Tab.~\ref{tab1}. We obtain that the spatial curvature is model-independently constrained to be  $\Omega_K=-0.140_{-0.158}^{+0.161}$. It is suggested that this is also well consistent with the spatially flat Universe. Moreover, compared with what obtained from the Union2.1 SN Ia, there is a subtle improvement in precision when the JLA SN Ia is considered.

\begin{table*}
	\centering
	\begin{tabular}{ccccccc}
	 \hline\hline
		~&Union2.1~SN~Ia&~and~&Cosmic~chronometers&~\\
		\hline
		$\Omega_K$~~~&~~~$\alpha$~~~&~~~$\beta$~~~&~~~$\delta$~~~&~~~$M_\mathrm{B}$ ~~~ \\
		\hline
		$-0.045^{+0.176}_{-0.172}$ &$0.120^{+0.009}_{-0.009}$ & $2.568^{+0.071}_{-0.070}$ & $0.034^{+0.032}_{-0.032}$  &  $-19.268^{+0.017}_{-0.017}$ \\
		\hline\hline
		~&JLA~SN~Ia&~and~&Cosmic~chronometers&~\\
		\hline
		$\Omega_K$~~~&~~~$\alpha$~~~&~~~$\beta$~~~&~~~$M_\mathrm{B}^1$~~~&~~~$\Delta_\mathrm{M}$ ~~~ \\
		\hline
		$-0.140^{+0.161}_{-0.158}$  & $0.140^{+0.009}_{-0.009}$ & $2.983^{+0.114}_{-0.112}$ & $-19.040^{+0.016}_{-0.016}$ & $-0.038^{+0.018}_{-0.018}$ \\
		\hline\hline
	\end{tabular}
	\caption{Summary of model-independent constraints on the spatial curvature and light-curve fitting parameters from SNe Ia and cosmic chronometers observations.}\label{tab1}
\end{table*}

\section{Conclusions}
In this paper, firstly, we reconstruct a function of Hubble parameter with respect to redshift, $H(z)$, with expansion rate measurements obtained from cosmic chronometers observations. The non-parametric smoothing method and the cutoff of redshift $z<1.2$ for observational data assure that the reconstructed function depends on neither cosmological nor stellar population synthesis models. Next, we obtain the comoving distance by directly solving the integral of this reconstructed function. We present the function of $H(z)$ smoothed from cosmic chronometers observations and the reconstructed comoving distance in unit of $c/H_0^{-1}$ in the left and middle panel of Fig.~\ref{fig1}, respectively. Furthermore, with the spatial curvature $\Omega_K$ taken into consideration, we can transform the comoving distance into the luminosity distance via Eq.~\ref{eq1}. The reconstructed distance modulus-redshift relations with differen $\Omega_K$ considered are shown in the right panel of Fig~\ref{fig1}. Clearly, these treatments are mathematical processes. Therefore, the curvature-dependent distance modulus reconstructed from  cosmic chronometers observations only depends on the assumption of  homogeneous and isotropic FLRW metric but has nothing to do with the matter-energy content of the Universe. More conventionally, luminosity distance is measured from SNe Ia observations. In this context, as shown in Eq.~\ref{eq2}, distance modulus is usually expressed as a linear combination of observed light-curve parameters ($m^*_\mathrm{B}$, $x_1$, and $c$). Coefficients in this expression ($\alpha$, $\beta$, and $\delta$), which need to be calibrated, are termed light-curve fitting parameters. In order to dodge the reliance on any assumptions of cosmological model, we directly confront the curvature-dependent distance (Eq.~\ref{eq1}) from cosmic chronometers observations with light-curve fitting parameters-dependent distance (Eq.~\ref{eq2}) from SNe Ia observations to obtain constraints on these undetermined coefficients. These cosmological-model-independent results are shown in Figs.~(\ref{fig2}, \ref{fig3}) and Tab.~\ref{tab1}. We find that the spatial curvature is constrained to be $\Omega_K=-0.045_{-0.172}^{+0.176}$ and $\Omega_K=-0.140_{-0.158}^{+0.161}$ when the Union2.1 and JLA SN Ia is used, respectively. It is suggested that these are in excellent agreement with the spatially flat Universe. Moreover, compared to the latest model-independent estimations of the spatial curvature with the distance sum rule~\citep{Rasanen2015}, results in our analysis are significantly improved in precision. This improvement might be very helpful to break degeneracies between the curvature and some other  important problems such as the evolution of the Universe and the nature of dark energy.

\section*{Acknowledgements}
This work was supported by the Ministry of Science and Technology National Basic Science Program (Project 973) under Grants No. 2014CB845806,  the Strategic Priority Research Program ``The Emergence of Cosmological Structure" of the Chinese Academy of Sciences (No. XDB09000000), the National Natural Science Foundation of China under Grants Nos. 11505008, 11603015, and 11373014, the China Postdoc Grants Nos. 2014T70043, and the Youth Scholars Program of Beijing Normal University.

\end{document}